\documentstyle[aps,prl,preprint]{revtex}
\begin{document}
\draft
\title{Inhibited Al diffusion and
growth roughening of Ga-coated Al (100)}
\author{Vincenzo Fiorentini, Donatella Fois, and Sabrina Oppo}
\address{Istituto Nazionale di Fisica della Materia - Dipartimento di 
Scienze Fisiche, Universit\`a di Cagliari, 
I-09124 Cagliari, Italy}
\date{Received 10 May 1996}
\maketitle 
\begin{abstract}
Ab initio calculations indicate
that the  ground state for  Ga adsorption on Al (100) is 
on-surface with local unit coverage.
On Ga-coated
Al (100), the bridge diffusion 
barrier for Al is large, but the Al$\rightarrow$Ga
{\it exchange barrier is zero}: the  ensuing
 incorporation of randomly deposited Al's into the Ga overlayer
realizes a percolation network, 
  efficiently recoated by Ga atoms.
Based on calculated energetics, we predict 
rough surface growth
  at all  temperatures; modeling the growth by
a random  deposition model 
with partial relaxation, we find a    
power-law divergent roughness  $w\sim t^{\,0.07\pm0.02}$.
\end{abstract}

The role played by  chemisorbed species
in modifying  energetics of
and adatom diffusion  on solid surfaces
 has recently come into focus in the field
of epitaxial metal-on-metal growth.
For example, the use of surfactants (agents inducing two-dimensional
growth) has become common
 in recent years for metal surfaces, a classic case  
being that of Sb on Ag (111) (widely studied 
experimentally \cite{sbagexp} and theoretically \cite{noi,altri}):
Sb adsorbates modify the nucleation behavior, island shapes, and ultimately
the growth mode of Ag (111) 
via a repulsive interaction with Ag adatoms
 \cite{noi,altri}.

The changes in adatom diffusion and growth 
mechanisms  that can be induced by
surface contaminants may however be quite different:
for instance, adsorbed species  may turn the growth mode
 to three-dimensional instead.
In this Letter we present one such ``reverse surfactant'':
a one-layer Ga coating inhibits Al diffusion on Al (100),
and  causes  growth roughening, independently of temperature (i.e.
 diffusion related) effects.

Our predictions are based on ab initio theoretical results
on  Ga adsorption and
Al/Ga coadsorption on Al (100). Ga wets Al 
(100), and the Ga monolayer adsorbed therewith on Al 
induces an anomalous diffusion mode  for deposited Al adatoms:
the barrier for Al$\rightarrow$Ga {\it exchange}  is zero. 
Thereby
 Al is embedded in the
adsorbed Ga monolayer and its 
diffusivity drops to zero, while Ga 
floats to the surface with  basically no barrier. 
The  random incorporation of  Al into the Ga coating layer
realizes, at submonolayer
coverage $\theta_{\rm Al}$, a percolation system with 
occupation probability $p = \theta_{\rm Al}$.
The  growth mode  of further layers
is disordered (roughening) at all temperatures,
due to the Ga-inhibited diffusion of Al adatoms and the
concurrent efficient Ga  recoating of the growing surface.
Using a simple  growth model we predict
a power-law time divergence of the  roughness, with exponent
$\beta\sim 0.07 \pm0.02$.
In a different direction, our results suggest
 Al as a candidate substrate for  the
growth of thin-films of Ga in the dense phase Ga-III
(unstable under normal conditions). Another
 reason of interest in this system is 
that liquid Ga embrittles Al, with
(100) as the fracture  plane\cite{rs2}.

Our ab initio total energy and force 
calculations are done within local density functional 
theory \cite{dft}, using a plane-waves basis with a cutoff of 16 Ryd,
 fully-separable norm-conserving pseudopotentials\cite{bhstm}, and
iterative diagonalization in a parallel
implementation \cite{rick}.
All geometries are fully optimized using Hellmann-Feynman forces.
Metallic behavior is accounted for via
a smeared occupation function  with 
a smearing of 0.1 eV, and
the first-order Methfessel-Paxton 
approximation to the $\delta$ function \cite{mp}. 
For all calculations,  we use downfolded special {\bf k}-point
 meshes equivalent to the  
Chadi-Cohen \cite{cc} 10 points for bulk Al.
These ingredients give a good descriptions of Al bulk and of the 
clean (100) surface \cite{bulkdat}.

\paragraph*{G\lowercase{a} adsorption on A\lowercase{l} (100) ---}

The adsorption of Ga on Al (100) was studied at coverages between 1
and 1/4 in the  fourfold-hollow on-surface and substitutional sites.
The adsorption energy is defined in the standard way (see e.g. 
Ref.\cite{noi}). The adsorbate chemical potential 
is chosen to be the   free-atom total energy 
\cite{cacca}.
 The calculated adsorption energies, given  in Table \ref{t1}, show
that substitutional adsorption is favored over on-surface 
 and subsurface sites by about 0.3 eV/atom
 at coverage $\theta_{\rm Ga}\leq$ 1/2 (i.e. whenever 
it is  meaningful): 
the adsorbate prefers to sit
 {\it  within} the surface layer, which    implies 
surface segregation and surface confinement of Ga on Al (100)
(compare with the analogous behavior of  Sb/Ag (111)
in Ref.\cite{noi}).

However,   on-surface adsorption
for  {\it unit}  Ga coverage is  
{\it energy-degenerate} with  substitutional adsorption. 
The  binding energy for on-surface Ga adsorption increases with
 increasing Ga coverage, and therefore there exists  an  {\it  effective
attraction between on-surface Ga adatoms}.
 The adatoms will tend to cluster 
and nucleate islands of local unit coverage
 even for $\theta_{\rm Ga} <$ 1.  
(This is opposite to the interadsorbate {\it repulsion} 
of Sb on Ag for all sites.)

Since thermal activation 
is needed (in the form of the  creation of surface vacancies) 
for the substitutional adsorption mode to be actuated, 
at all temperatures sufficient to activate Ga diffusion on Al but
not high enough to create surface vacancies
the adsorption ground state will be the  on-surface site. 
 To estimate the  temperature at which  substitutional
adsorption sets in, one must evaluate the relevant activation
 barrier. While it is relatively easy to calculate
the formation {\it energy} of a surface vacancy (we
 find it to be 0.69 eV/vacancy at $\theta_{\rm vac}$= 1/4), it is extremely 
difficult to accurately estimate the {\it barrier}
against
its formation. An  intermediate configuration which is expected 
to be unavoidable at some stage of the substitutional-Ga 
adlayer formation process is
the adatom-vacancy  Frenkel pair (a vacancy with the removed Al atom on
the surface): its formation energy may then provide an estimate
of the barrier in question.  We calculated the pair formation energy 
to be 1.01 eV/defect at coverage 1/4. Assuming this energy as  an
estimate of the activation barrier, we infer that a
 temperature of T$_{\rm act} \sim$ 
450 K is needed to activate  one event per second
(estimated as in Ref. \cite{rs},
assuming a Boltzmann prefactor of 10$^{-2}$ cm$^2$/s,
and a diffusion length of 50 bohr, which  
the ejected adatom is supposed to travel
 to get to a kink site; this corresponds 
 to a reasonable step density of 10$^{13}$/cm$^2$).
According to this estimate,  on-surface Ga adsorption will dominate 
at room temperature \cite{rs3}. 

\paragraph*{Anomalous A\lowercase{l} diffusion on G\lowercase{a}:A\lowercase{l}
(100), and its consequences for the growth of A\lowercase{l} ---}

As surface diffusion is a crucial factor in epitaxial growth, and
given the stability of a Ga monolayer on Al (100),
it is natural to consider the way in which this
surface coating affects the motion of Al adatoms.
The bridge and exchange  diffusion paths  were 
 studied for Al on Ga-contaminated Al 
(100) for $\theta_{\rm Ga}$=1 
and  adatom coverage of $\theta_{\rm Al}$=1/9
 \cite{nota_cut}. 
The barrier for the bridge transition state is 0.63 eV. 
However, for the transition state of the
exchange reaction \cite{rs,feib2}
between Al and Ga, we find  a barrier which is
zero within computational accuracy.
The final state of the exchange process (Ga on-surface and Al embedded 
into the Ga monolayer) is favored by 0.29 
eV with respect to the initial state
(Al on-surface on the intact Ga monolayer). The adatom is thus 
incorporated within the Ga monolayer
with no need for  thermal activation, and a Ga atom
concurrently pops up onto the surface. At low temperature,
the Al$\rightarrow$Ga exchange process is essentially irreversible, and
the  mobility of the Al atom embedded into the Ga adlayer is expected 
to be negligible: thus the Al adatom becomes immobile and its
 diffusivity drops to 
zero.
While unexpected in general, a  drop
 in diffusivity
as the diffusion barrier gets lower
is  justified in this case by the  asymmetry (Al-Ga) 
of the process \cite{nota_dif}.

The exchange process  amounts to an individual
segregation event for Ga: the absence of a barrier implies
that  Ga will  efficiently float to the surface 
despite the moderate energy gain obtained therewith.  
The calculated adsorption energetics (Table \ref{t1}) 
for the substitutional and sublayer sites indicates
that segregation of Ga will be 
complete even at higher $\theta_{\rm Al}$.
This is further confirmed (in addition to  the above-mentioned 0.3 
eV/atom  gain upon Al$\rightarrow$Ga exchange at $\theta_{\rm Ga}$=1 and
 $\theta_{\rm Al}$=1/9)  by  ({\it a}) 
a 0.3 eV/atom gain forming an on-surface Ga monolayer 
on perfect Al (100) from a mixed Ga$_{0.5}$-Al$_{0.5}$ 
surface alloy bilayer, 
({\it b}) a 0.4 eV/atom gain upon Al$\rightarrow$Ga exchange 
at $\theta_{\rm Ga}$=1/4 and $\theta_{\rm Al}$=1/4 \cite{rs3}.
Since Ga efficiently recoats
 the surface (see below),  the Al$\rightarrow$Ga exchange barrier 
should not depend appreciably on Al coverage, so Ga segregation is
always very efficient.

The deposited Al adatoms exchange with Ga atoms from the
wetting adlayer with essentially no diffusion prior to incorporation,
a diffusion-unbiased hit-and-stick process.
This process leads to the formation of a high-areal-density array of 
small (mostly monoatomic at low coverage), randomly-distributed Al clusters 
embedded within the Ga surface layer. 
If we formally view the  Ga adlayer as an 
empty square lattice, upon deposition of
a fraction $\theta_{\rm Al}$  of an Al submonolayer the  sites of this
 lattice are randomly occupied by Al adatoms
with a probability $p$\,=\,$\theta_{\rm Al}$. 
A percolation system is thus realized whose
site occupation probability is  equal to (and  tunable via) the Al coverage.
Using the standard results for site percolation on a two-dimensional 
square lattice \cite{stau}, below the 
percolation-threshold coverage $\theta_{\rm Al}^c\equiv p_c = 0.593$ the 
Al clusters are finite and have a  fractal dimension D$\sim$1.2,
becoming D$\sim$1.9 at $p_c$. Above $p_c$, the
percolating clusters have Hausdorff dimension 2.
This  system may open interesting new possibilities,
as for instance the direct study of percolation clusters
and of tracer diffusion on a fractal cluster.

The growth of  further layers is expected to proceed in a disordered
fashion and roughening occurs with  essentially no 
temperature dependence.
Let us first assume that  the Ga atoms kicked off the wetting layer are
removed from the surface. While Al adatoms 
diffuse normally on previously-created Al areas, they are
embedded instantaneously in the surface layer
 upon entering  Ga-covered areas, which  effectively 
act as drains for Al adatoms. This diffusion mode 
is equivalent to downstep diffusion with zero step barrier.
Two-dimensional growth of	
the first layer follows immediately.
Of course, however, the Ga atoms substituted for by Al adatoms 
 remain on the surface. These Ga atoms diffuse 
rapidly on the Ga-covered areas,
they are attracted onto the  previously created Al 
regions, and recoat them. This is because  ({\it a}) 
Ga is less bound to Ga/Al(100) than to
Al(100) by 0.2 eV, and ({\it b}) the
bridge  diffusion barrier for
 Ga on Ga:Al(100) is as low as 0.01 eV  at $\theta_{\rm Ga}$=1
\cite{rs2}, which we expect to hold qualitatively at lower coverage too,
given the lower binding energy to the substrate 
(point ({\it a})). Since  Al adatoms do not diffuse and
 deposition is random, further incoming Al's
 will be embedded with equal probability
within the newly created Ga overlayers (``second-level'' Ga) as 
on the pristine Ga layer. New  Al clusters
will ``nucleate'' (by Al-Ga exchange) on 
top of previously embedded Al clusters as well 
as on pristine Ga  areas. This process is analogous to 
overgrowth on an incomplete layer,  so that 
three-dimensional growth ensues. 
This is all the more likely when the shape of the Al clusters 
is reasonably regular, i.e. for  $\theta_{\rm Al}$ above
the percolation threshold $p_c$ (where clusters have D = 2 and a
 relatively  regular  shape).
In a sentence, the surface is coated 
with a diffusion-inhibiting segregating agent, causing
growth roughening. For the latter to 
occur, the only  condition is that
Ga diffusion be activated so as to allow recoating of the Al clusters.
As discussed above, this is a rather weak condition indeed; also, it is
in fact immaterial at low temperature,
because  the substituted Ga will hardly leave at all the 
vicinity of its original site.

The hit-and-stick deposition mode emerging from the 
results discussed above leads to Poisson growth, with
a roughness  diverging in time as   $t^{1/2}$.
However, one should  account for the fact that
below and around the
percolation threshold  $\theta_{\rm Al}^c=p_c$  
the surface-embedded Al clusters are fractals, and their shape (and that of
the  ``second level'' Ga recoating areas) is highly irregular. Thus 
 diffusion of Al down the highly ramified second-level Ga clusters 
may take place before exchange does. In view of this fact 
we model the growth of the system by random deposition with
{\it partial relaxation}: adatoms  stick where they land, except on the
thinnest (one-atom wide) second-level Ga islands and  at island edges:
in practice, they  move from the landing site to a
nearest-neighbor site if the latter is lower in height, and stick there.
We  analyzed the   statistical properties\cite{barab} of our
 partial 
relaxation model by 2+1-dimensional lattice simulations:
the results indicate a  roughness 
diverging in time as a power law, $ w \sim t^{\beta}$, with
 $\beta\simeq 0.07 \pm 0.02$ \cite{nota_exp}. This 
 agrees with the finiteness of the range of lateral correlations.
The interesting point is that since the model seems 
fairly realistic for this system, it could possibly be 
experimentally tested. A more detailed statistical analysis
  will be given  elsewhere.

\paragraph*{A\lowercase{l} as substrate for G\lowercase{a} growth ---}

The general features of the Ga/Al system 
correlate with the size
differences of Ga and Al in dense bulk phases.
Ga is slightly larger than Al, and the two metals do not form 
high-density  (nor other) alloys. This is compatible with 
both surface segregation with a moderate energy gain, and  
with the small in-plane lattice mismatch $m =
[a_{\rm Ga}/a_{\rm Al}$ -- 1] of Ga to 
Al. Fcc Ga (for which $m \sim$ 1.5 \%) distorts 
spontaneously into face-centered  tetragonal Ga-III \cite{bern},
whose mismatch to  Al (100) is $m \sim$ --1.7 \%,
These values are small enough  to make  pseudomorphic 
 wetting of Ga on Al (100) favorable; so, geometry-wise,
 the stability of the Ga monolayer adsorbed on Al (100)
seems  just natural.

Our calculated surface energy of  Ga:Al(100) at 
$\theta_{\rm Ga}=1$ is 0.12 eV/atom =24 meV/\AA$^2$ 
assuming  the bulk energy of 
Ga-III  (energy-degenerate with fcc Ga) as Ga chemical potential.
This is much lower than the
calculated \cite{bern} surface energy of
Ga-III (0.56 eV/atom=74 meV/\AA$^2$):
this  gain in surface energy suggests the
use of Al (100) as substrate for  thin-film growth of Ga-III.
Assuming volume conservation, Ga-III  accomodates
its small in-plane lattice mismatch 
to fcc Al with a mere  3.5\% expansion along the growth direction, so that
 pseudomorphic growth may be sustained up to appreciable 
thicknesses. It was shown recently \cite{bern} that the 
(100) surface of the stable $\alpha$-Ga phase reconstructs so
as to expose a   Ga-III--like surface bilayer,
so that the {\it surface} structure of Al-epitaxial Ga
should remain Ga-III--like at all film 
thicknesses. However, dimerization of the 
Ga film  into the $\alpha$-phase structure above 
some critical thickness cannot be 
ruled out. These issues 
will be discussed elsewhere.

\paragraph*{Low barrier for exchange: an estimate ---} 
The low cost of the
Al$\rightarrow$Ga exchange may be explained
 similarly to 
Al-Al exchange on Al (100) (each atom involved
 is always at least three-fold coordinated, while 
at a  bridge saddle point the adatom is only 
two-fold coordinated \cite{feib2}).
The bond lengths and adsorbate heigths at the 
exchange saddle point (see Table \ref{t2}) support this
view: the exchanging-atoms dimer burrows about 4\% 
deeper into the surface layer than in the
Al-Al exchange on Al(100); also,  not only does the exchanging Al 
adatom form short bonds to the three neighboring top-layer Ga atoms,
but it is also {\it already bound to a second-layer Al},
forming a bond  2\% shorter than  bulk ones (about the same 
holds for the exchanging Ga). The overall number of bonds
is 9 in the  Al$\rightarrow$Ga exchange, against 6 of
 the Al$\rightarrow$Al exchange on clean Al. 
The initial and final states
have a total of  12 bonds in  both cases. A bond-cutting 
model allowing for bond saturation
at high coordination \cite{mhs}, gives 
 a gain of $\sim$0.4 eV for a reduction of the number of cut bonds 
from 6 to 3,  matching
 the $\sim$ 0.3 eV difference of exchange barriers for clean Al 
 \cite{rs,feib2} and  Ga:Al(100).
Coordination models often give inaccurate barrier values for 
 different diffusion processes (the model above
 predicts a barrier of 0.9 eV for both exchange and
bridge diffusion, the  calculated \cite{rs,feib2}
 values  being  0.3 eV and 0.6 eV),
but one expects error cancellation 
when estimating a  difference between barriers 
of  two very similar
processes (as done here \cite{pippa}), so the 
 agreement is probably not accidental.

\paragraph*{Summary ---}  

Ga adsorption on Al (100) has
a degenerate ground state: on-surface adsorption  
of local coverage unity at all 
temperatures, and 
thermally activated (above T $\sim$ 450 K)
substitutional adsorption at low coverage.
The diffusion barrier for Al on Ga-coated Al (100) ($\theta_{\rm 
Ga}$=1) is found to be large
 for bridge diffusion, but null for the Al$\rightarrow$Ga 
exchange process, in which Al gets 
locked into the surface layer while Ga jumps onto the surface. 
Al is  incorporated immediately upon deposition,
while Ga efficiently segregates to the surface. 
At submonolayer Al coverage, Al incorporation in the Ga adlayer
effectively realizes a percolation system, 
which is recoated by Ga as the growth proceeds.
The surface grows rough, with a power-law
diverging roughness $w\sim t^{0.07\pm0.02}$; this 
agrees qualitatively with the 
 strong diffusion-inhibiting action of the
 Ga coating.
Finally, we suggest that Al may  be a good substrate
 for thin-film growth of metastable
Ga-III.

\paragraph*{Acknowledgements ---}
We thank  Roland Stumpf and Alessio 
Filippetti for discussions, and Riccardo Valente and 
Stefano Baroni for providing their  parallel code.
 CRS4 Cagliari provided  IBM SP2 computing time 
  within a collaboration with 
the University  of Cagliari.


%
\begin{table}
\begin{tabular}{l|ccc} 
$\theta$ &  1/4 &  1/2 &  1 \\ \tableline
E$_{\rm ad}^{\rm subst}$ &  3.83 &  3.82 &  --- \\ 
relax &  +3.0    &  +4.0    &  --- \\ 
W &  4.43 &  4.04 &   --- \\ \tableline
$E_{\rm ad}^{\rm on-srf}$ &  3.52 &  3.48 &  3.80 \\ 
relax &  --7.0    &  --11.6    &  +4.9    \\ 
W &  4.53 &  4.49 &  4.57 \\ \tableline
$E_{\rm ad}^{\rm subsurf}$ &  3.59 &  3.57 &  3.57\\
relax &  --1.1  &  +1.3  &  --0.1  \\ 
W &  4.39 &  4.18 &   4.41 \\
\end{tabular}
\vspace{0.25cm}
\caption{ Adsorption energies (eV),
adatom relaxations (percentage of interlayer distance), and
work function (eV) 
for Ga on Al (100) in the on-surface, substitutional, and subsurface
sites (see text for details).}
\label{t1}
\end{table}
\begin{table}
\begin{tabular}{l|cc} 
Atom pair & This work & Ref.\cite{rs}, Al:Al(100) \\
\tableline
h$_{\rm Al_{\rm ex}}$ & 1.08 & 1.40\\
h$_{\rm Ga_{\rm ex}}$ &1.15 & \\
dimer length & 4.77 & 4.84\\
Al$_{\rm \,ex}$-Ga$_{\rm \,sub}$ & 4.70 & 4.75\\
Al$_{\rm \,ex}$-Al$_{\rm \,sub}$ & 5.20&\\
Ga$_{\rm \,ex}$-Ga$_{\rm \,sub}$ & 4.57&\\
Ga$_{\rm \,ex}$-Al$_{\rm \,sub}$ & 5.21& \\
\tableline
\end{tabular}
\vspace{0.25cm}
\caption{Bond lengths and heights (bohr) for  Al$\rightarrow$Ga exchange
on Ga:Al(100).  Bond length in bulk Al: 5.30 bohr.
  X$_{\rm \,ex}$ are exchanging atoms,
  X$_{\rm \,sub}$  substrate. Right column: values for 
Al-Al exchange on Al (100) (only Al's involved).}
\label{t2}
\end{table}

\begin{references}
\bibitem{sbagexp}
H. A. van der Vegt {\it et al.}, Phys. Rev. Lett. {\bf 68}, 3335 (1992);
J. Vrijmoeth {\it et al.}, {\it ibid.} 
{\bf 72}, 3843 (1994); N. Memmel and E. Bertel, {\it ibid.} {\bf 75}, 
485 (1995); K. Bromann {\it et al.}, {\it ibid.}, 677;
J. A. Meyer  {\it et al.}, Phys. Rev. B {\bf 51}, 14790 (1995).
%
\bibitem{noi}
S. Oppo, V. Fiorentini, and M. Scheffler, 
Phys. Rev. Lett. {\bf 71}, 
2437 (1993); MRS Proc. {\bf 317}, 323 (1994); 
V. Fiorentini, S. Oppo, and M. Scheffler, Appl. Phys. A {\bf 60}, 399
(1995);
 M. Scheffler, V. Fiorentini, and S. Oppo, in  {\it 
 Surface Science},  edited by R. McDonald, A. Taglauer, and K. Wandelt
(Springer, New York, in press).
\bibitem{altri}
J. Tersoff, A. W. Denier van der Gon, and R. M. Tromp, 
Phys. Rev. Lett. {\bf 72}, 266 (1994);
S. Liu {\it et al.}, {\it ibid.} {\bf 74}, 4495 (1995).
%
\bibitem{rs2}
R. Stumpf and P. J. Feibelman, unpublished.
\bibitem{dft}
 R. Dreizler and E. K. U. Gross, {\it Density functional theory}, (Springer,
Berlin, 1990). Exchange-correlation energy  by
D. M. Ceperley and B. J. Alder, Phys. Rev. Lett. {\bf 45}, 566 (1980)
as  parametrized by J. P. Perdew and A.  Zunger,
Phys. Rev. B {\bf 23}, 5048 (1981).
\bibitem{bhstm}
L. Kleinman and D. Bylander, Phys. Rev. Lett. {\bf 48}, 1425 (1982);
U. von Barth and R. Car, unpublished.
%
\bibitem{rick}
R. Valente and S. Baroni, to be published.
%
\bibitem{mp}
M. Methfessel and A. P. Paxton, Phys. Rev. B {\bf 40}, 3616 (1989).
This choice eliminates deviations of the calculated total energy from
its correct zero-smearing value up to and including third order 
in the smearing. The Hellmann-Feynman forces also come out automatically 
correct. See  S. de Gironcoli, Phys. Rev. B {\bf 51}, 6773 (1995) for an 
up-to-date discussion.
%
\bibitem{cc}
D.J.
 Chadi and M.L. 
Cohen, 
Phys. Rev. B {\bf 7}, 5747 (1973).
%
\bibitem{bulkdat}
For Al bulk, $a_0^{\rm th} = 7.50$ bohr
(exp.: 7.65) and $B = 780 $ Kbar (exp.: 783). The theoretical lattice 
constant is used in all calculations.
For Al (100) the surface  energy is 0.071 eV/\AA$^2$,
 and the surface stress is 0.099 eV/\AA$^2$.
\bibitem{cacca}
 Calculated 
in a large supercell at the same cutoff;
spin polarization not included. To obtain the surface 
energy referred to a bulk
reservoir, one uses an appropriate  bulk energy as chemical potential
of the adsorbate.
%
\bibitem{rs}
R. Stumpf, {\it Gesamtenergierechnungen zu Adsorption, Diffusion und
Wachstum auf Al-Oberfl\"achen} (K\"oster, Berlin, 1993);
R. Stumpf and M. Scheffler, 
Phys. Rev. Lett. {\bf 72}, 254 (1994); Phys. Rev. B {\bf 53}, 4958 (1996).
%
\bibitem{rs3}
Stumpf and Feibelman \cite{rs2} report that an Al adatom 
 on Al (100) with substitutional Ga 
at $\theta_{\rm Al}$=$\theta_{\rm Ga}$=1/4 
(a Frenkel pair with a Ga atom embedded in the vacancy)
is 0.4 eV higher in energy than the 
same system in which the adatom  is  at a bulk site. Assuming {\it this}
value as a lower bound for  the barrier, 
the activation temperature for one event/second would be
about 185 K. However  this is probably an underestimate,
as the configuration in question is likely to
require vacancy formation to begin with.
\bibitem{nota_cut}
For these calculations,
the cutoff was reduced to 8 Ryd, which was found
to leave  the bridge diffusion for Al on Al (100) and Al 
on Ga:Al(100) unaltered 
to within 3 \%.
%
\bibitem{feib2}
P. J. Feibelman, Phys. Rev. Lett. {\bf 65}, 729 (1990).
%
\bibitem{nota_dif}
Indeed the symmetric Al-Al exchange diffusion on clean Al 
(100), having a lower barrier than bridge diffusion, causes 
an  enhancement  of Al 
diffusivity \cite{rs,feib2}.
\bibitem{stau}
D. Stauffer, {\it Introduction to percolation theory}
(Taylor and Francis, London, 1985).
\bibitem{barab}
A. L. Barabasi and H. E. Stanley, {\it Fractal Concepts in 
Surface Growth} (Cambridge U.P., Cambridge, 1995).
\bibitem{nota_exp}
As the divergence is weak, extracting  
the growth  exponent (the slope of log $w$ vs 
 log $t$) is somewhat tricky. In the whole span
from deposition start to  size saturation,
the data may also be crudely fit by
log $ w\sim$ 0.09 log $t$ -- 0.01 (log $t$)$^2$.
This is close to a simple power law, with a
slighty different exponent. Roughness evolution data 
were obtained as averages  over 100 runs of 3000 steps on 
a 200$\times$200 lattice.
\bibitem{bern}
M. Bernasconi, G. L. Chiarotti, and E. Tosatti, 
 Phys. Rev. Lett. {\bf 70} 3295 (1993); Phys. Rev B {\bf 52}, 9988 
(1995); {\it ibid.}, 9999.
%
\bibitem{mhs}
 M. Methfessel, D. Hennig, and M. Scheffler, Phys. Rev. B {\bf 46}, 
4816 (1992); Appl. Phys. A {\bf 55}, 442 (1992).
\bibitem{pippa}
We  assumed the same energy-coordination 
relation  for all mixed bonds. This seems 
admissible for a crude estimate.
%
\end{references}
\end{document}